\documentclass[letter]{aa}
\usepackage{graphicx}
\usepackage{natbib}
\usepackage{epsfig}
\usepackage{subfigure}
\usepackage{multirow}
\usepackage{txfonts}
\begin{document}
   \title{A fresh look at the seismic spectrum of HD49933: \\ analysis of 180 days of CoRoT photometry\thanks{The CoRoT space mission, launched on 2006 December 27, was developed and is operated by the CNES, with participation of the Science Programs of ESA, ESA's RSSD, Austria, Belgium, Brazil, Germany and Spain.}}
   
 \titlerunning{The solar-like oscillations of HD49933}


   \author{O. Benomar\inst{1} \and F. Baudin\inst{1} \and T.L. Campante\inst{2,3}   
          \and W.J. Chaplin\inst{4} \and R.A. Garc\'ia\inst{5} \and P. Gaulme\inst{1} \and \\
          T. Toutain\inst{6} \and G.A. Verner\inst{7} \and
          T. Appourchaux\inst{1} \and J. Ballot\inst{8} \and C. Barban\inst{9}
          \and Y. Elsworth\inst{4} \and S. Mathur\inst{10} \and B. Mosser\inst{9} \and \\
          C. R\'egulo\inst{11,12} \and
          I.W. Roxburgh\inst{7} \and M. Auvergne\inst{9} \and A. Baglin\inst{9}
          \and C. Catala\inst{9} \and E. Michel\inst{9} \and R. Samadi\inst{9}
          }

   \institute{Institut d'Astrophysique Spatiale, CNRS, Universit\'e Paris XI, 91405 Orsay, France
   	\and Danish AsteroSeismology Centre (DASC), Department of Physics
	 and Astronomy, University of Aarhus, 8000 Aarhus C, Denmark
	\and Centro de Astrof\'isica da Universidade do Porto, Rua das Estrelas, 4150-762 Porto, Portugal
	\and School of Physics and Astronomy, University of Birmingham,
	 Edgbaston, Birmingham B15 2TT, UK
	\and Laboratoire AIM, CEA/DSM-CNRS, Universit\'e Paris Diderot,
	IRFU/SAp, Centre de Saclay, 91191Gif-sur-Yvette, France
	\and Center for Information Technology, University of Oslo, PO.Box 1059 Blindern,N-0316 Oslo, Norway
	\and Astronomy Unit, Queen Mary, University of London Mile End Road, 
	London E1 4NS, UK
	\and Laboratoire d'Astrophysique de Toulouse-Tarbes, Universit\'e de Toulouse, CNRS, 31400 Toulouse, France
	\and LESIA, UMR8109, Universit\'e Pierre et Marie Curie, Universit\'e
	Denis Diderot, Observatoire de Paris, 92195 Meudon, France
	\and Indian Institute of Astrophysics, Koramangala, Bangalore 560034,
	India
	\and Instituto de Astrof\'isica de Canarias, 38205, La Laguna,
	Tenerife, Spain
	\and Universidad de La Laguna, Dpto de Astrof\'isica, 38206, La Laguna, Tenerife, Spain
             }

   \date{Received  12 August 2009 / accepted 15 October 2009}

 
  \abstract
  {Solar-like oscillations have now been observed in several stars,
  thanks to ground-based spectroscopic observations and space-borne
  photometry. CoRoT, which has been in orbit since December 2006, has
  observed the star HD49933 twice. The oscillation spectrum of this
  star has proven difficult to interpret. }
   {Thanks to a new timeseries provided by CoRoT, we aim to provide a
   robust description of the oscillations in HD49933, i.e., to
   identify the degrees of the observed modes, and to measure mode
   frequencies, widths, amplitudes and the average rotational
   splitting.}
   {Several methods were used to model the Fourier spectrum: Maximum
   Likelihood Estimators and Bayesian analysis using Markov Chain
   Monte-Carlo techniques.}
   {The different methods yield consistent result, and allow us to
   make a robust identification of the modes and to extract precise
   mode parameters.  Only the rotational splitting remains difficult
   to estimate precisely, but is clearly relatively large (several
   $\mu$Hz in size).}
   {}

   \keywords{stars : oscillations
               }

   \maketitle
%

 \section{Introduction}
 \label{sec:intro}

Stars are now the objects of seismic studies after decades of similar
studies for the Sun, thanks to the advent of space-borne photometric
observations (e.g., MOST, CoRoT and Kepler) and extremely precise
ground-based spectroscopic observations \citep[for a complete review, see for e.g.][]{Aerts2008SoPH}.
This applies in particular to
stars presenting solar-like p modes (acoustic oscillations
stochastically excited by convection), with CoRoT observations of such
stars showing clearly individual peaks in the Fourier spectra
\citep[e.g.][]{michel08}. Among these stars, HD49933 has already been
the target of asteroseismic campaigns. It was first observed
spectroscopically from the ground for 10 nights \citep{mosser05}. In
2007, a first photometric time series of 60 days was collected by
CoRoT, followed by a new long run of 137 days in 2008. HD49933 is a F5
main sequence star with an apparent visual magnitude $m_V = 5.77$. It
is hotter than the Sun \citep[$T_{\rm eff}$=6780\,K or $T_{\rm
eff}$=6500\,K,][]{bruntt08,bruntt09,ryabchi09} with an estimated mass of $\sim
1.2M_{\odot}$ \citep{mosser05} and an estimated radius of $1.34 \pm
0.06\,R_{\odot}$ \citep{thevenin06}. The surface rotation ($v\sin i$)
was determined to be around 10\,km\,s$^{-1}$
\citep{mosser05,solano05}. The surface rotation period has also been
measured at $\sim$\,3.4 days, using the 60-day CoRoT timeseries
\citep{appourchaux08,deheuvels2008}, from the signatures of photospheric
transiting active regions (e.g., spots) which give rise to a clear
peak in the very low-frequency part of the Fourier spectrum.

The seismic interpretation of HD49933 has proven to be very
difficult. \citet{mosser05} could not isolate individual p modes in
the Fourier spectrum of observed line-of-sight velocities but were
able to find a regular pattern in the spectrum, which is the signature
of the large frequency separation between modes of same degree $l$
(but increasing radial order $n$). The first CoRoT time-series was
analysed by \citet{appourchaux08}, and these data clearly show
individual p-mode peaks in the Fourier spectrum. However, the peaks
show large widths, making the interpretation less than
straightforward: a given peak could be interpreted as being a closely
spaced pair of $l$=0 and $l$=2 modes, or a single (but rotationally
split) $l$=1 mode. Based on the modeling of the spectrum using a
Maximum Likelihood Estimator fitting method, \citet{appourchaux08}
chose one of these two possible interpretations (hereafter called
model A) based on the highest {\em likelihood} of each model. This first time-series was the object of other studies. \cite{appourchaux2009a} put into perspective the results of \citet{appourchaux08}, showing that the likelihood ratio test does not give the probability of the hypothesis given the data, but only the significance of the data given the hypothesis. \cite{benomar09}, who applied a Bayesian analysis to the same time series, could not definitely favour one interpretation (model A) over the
alternate (model B), based on the {\em whole} probability distribution of each model. \citet{gruberbauer2009a}, using a Bayesian approach too, also consider the identification ambiguous. \citet{gaulme2009} used a simpler Bayesian approach (Maximum A Posteriori, or MAP, approach). The most probable model they found corresponds to the same identification as \citet{appourchaux08}. More recently, \citet{mosser09b} proposed an empirical method to determine the identification of the modes. Its application considers the model B as the more likely when using the two datasets used here.

It should be noted that the case of HD49933 is quite different from the solar case: solar modes are very narrow in comparison. Their widths (1\,$\mu$Hz for the modes with the
highest amplitudes) are much smaller than the small-frequency
separations between $l$=0 modes of order $n$ and the neighbouring
$l$=2 modes of order $n$-1 (being typically around 10\,$\mu$Hz for the
Sun). Moreover, the star inclination angle, if small, tends to attenuate the visibility of mode components with azimutal order $m\ne0$, making the mode identification even more difficult.

\begin{figure*}
\begin{center}
\includegraphics[angle=90,width=18.4cm,height=6.8cm]{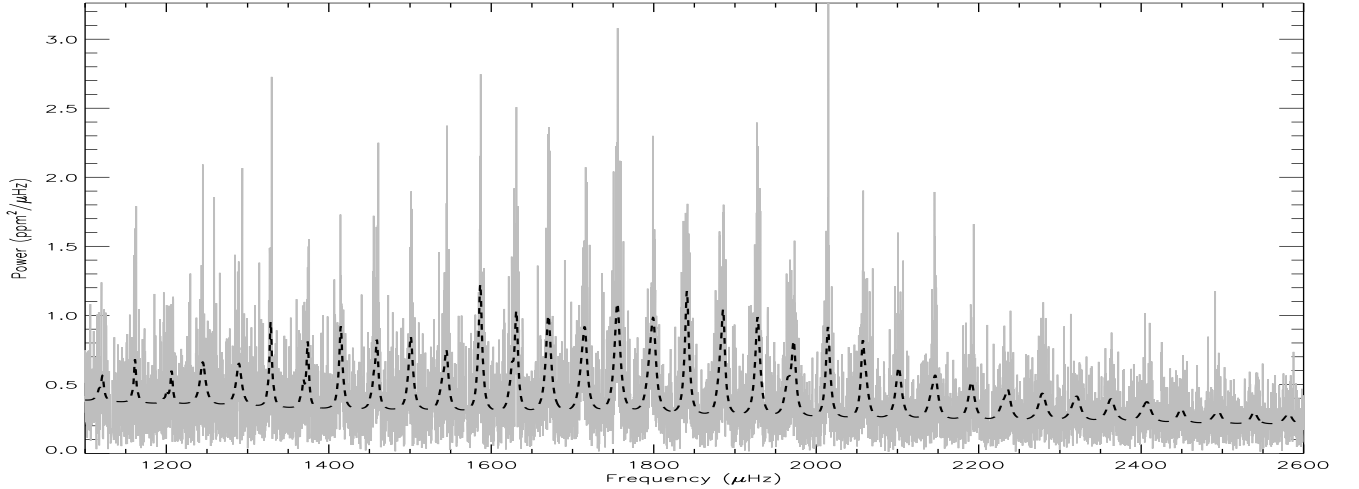}
\end{center}
\caption{Mean power spectrum using 3 time series of 60 days (solid grey line), and the fitted model (dashed line).}
\label{fig:spec}
\end{figure*}

\begin{figure*}
\begin{center}
   \subfigure{\includegraphics[angle=90,width=8.5cm]{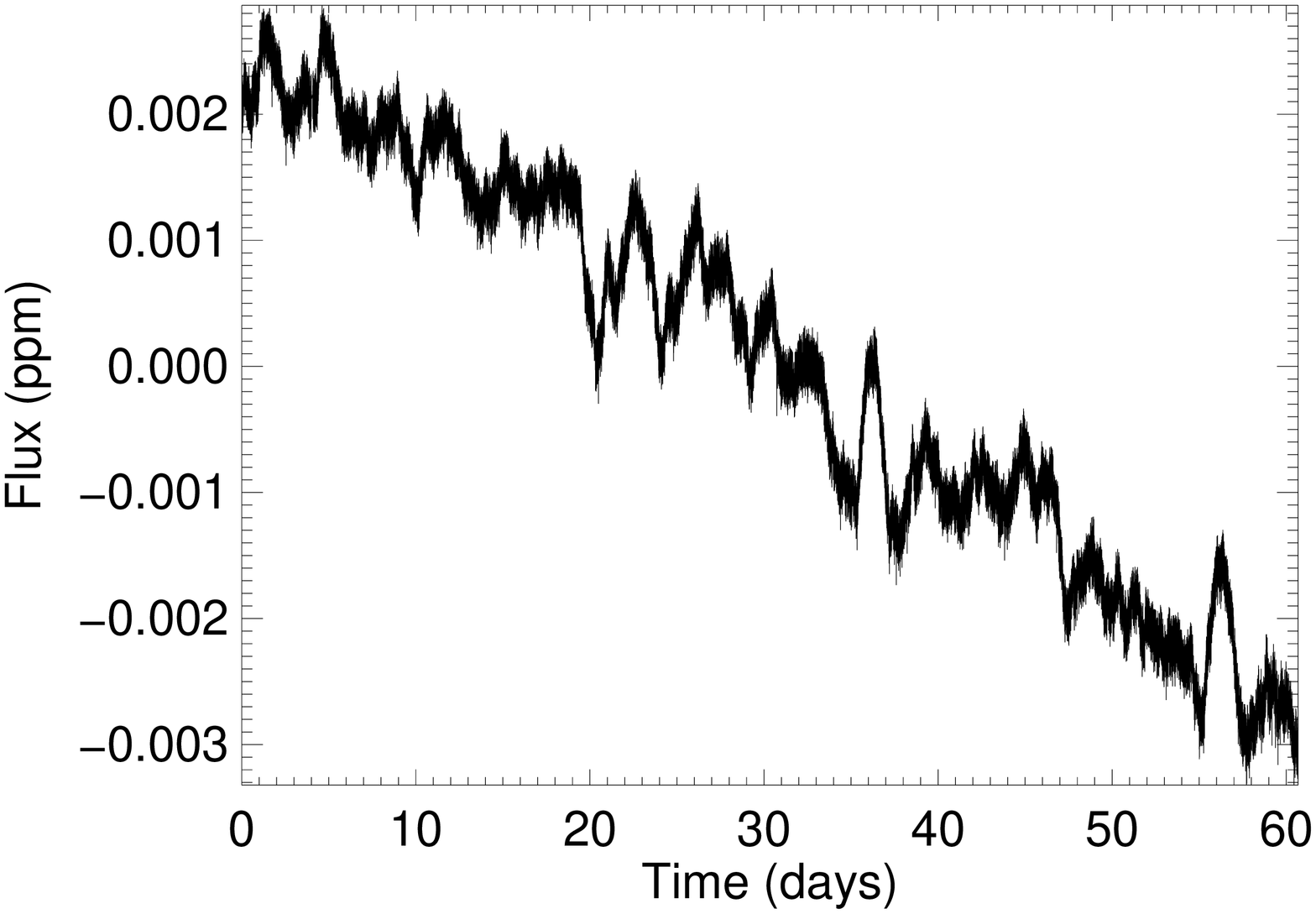}}
   \subfigure{\includegraphics[angle=90,width=8.5cm]{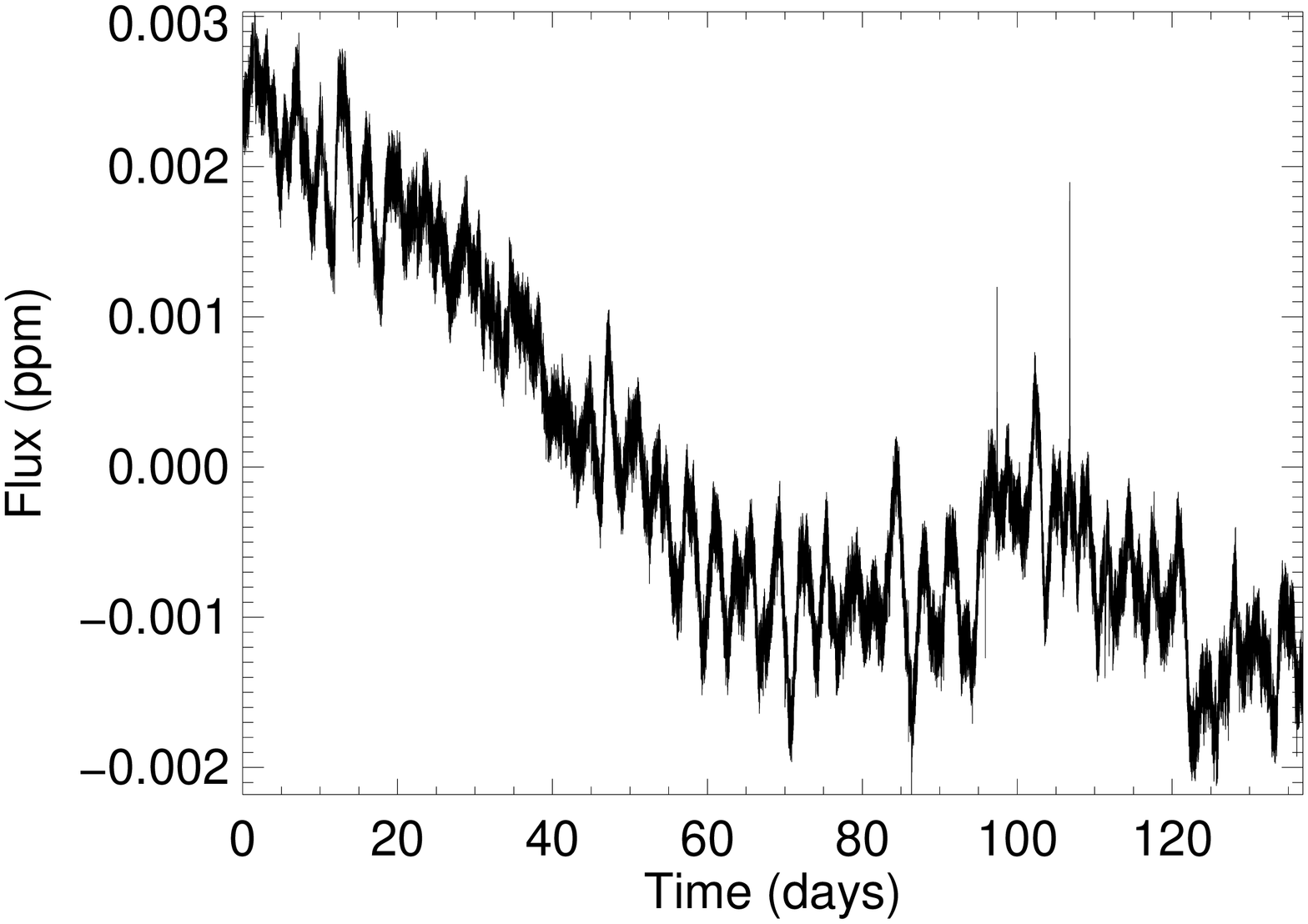}}
\end{center}
\caption{Lightcurves of HD49933 for the 60-day long initial run (left) and the 137-day long run (right)}
\label{fig:LC}
\end{figure*}

\begin{figure}
\begin{center}
\includegraphics[angle=90,height=6cm,width=10cm]{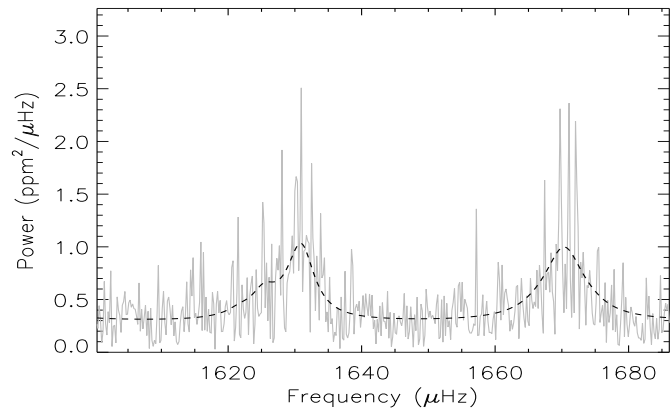}
\end{center}
\caption{Detail of the spectrum (solid grey line), and of the fitted model (dashed line) with a pair $l=0/2$ on the left and an $l=1$ on ther right.}
\label{fig:zoomspec}
\end{figure}

Here, we use the new long-run CoRoT observations of HD49933,
together with the original shorter run timeseries (Fig.\,\ref{fig:LC}), in order to properly describe the acoustic oscillations of the star clearly visible in the Fourier spectrum (Fig.\,\ref{fig:spec}, Fig.\,\ref{fig:diag_ech0} and Fig.\,\ref{fig:zoomspec}).

 \section{Methodology}
 \label{sec:method}

The time series used here were extracted in the same way as in
\citet{appourchaux08}. The gaps in the series represent slightly less than 10\% and were filled by linear interpolation.
In order to provide a robust seismic interpretation of the CoRoT
timeseries for HD49933, several analyses were performed, using
different methods (Maximum Likelihood Estimators, or MLE; and various
Bayesian analyses); or the same method applied in an independent
manner. These methods were already used in the same way to analyze the initial run of 60 days \citep{appourchaux08,benomar09}. The ``fitters'' of \cite{appourchaux08} and some additional fitters sought to find a best-fitting model
spectrum. The model has several contributions. There is a contribution
from the background, which includes signatures of convection and possibly
phenomena with longer time scales (e.g. those related to the stellar
activity) and contributions from the individual p modes. Each mode is
described by a set of parameters: a central frequency, a width and a
height.  A single height and width was fitted to each $l=0$/2 pair and
the closest $l=1$ mode (in frequency). The relative heights of the
$l$=0, 1 and 2 modes took fixed values, which were assumed to be
independent of frequency. A single rotational frequency splitting
parameter was fitted to \emph{all} non-radial modes. The stellar
inclination, which governs the relative heights of the different $m$
components for a given ($n$,$l$) mode, was also fitted as a single,
global parameter. The observed spectrum used by the different fitters
was an averaged spectrum (frequency resolution 0.19\,$\mu$Hz) made
from three timeseries of the same duration, which came from the two
CoRoT runs: the first 60-day run, IRa01, and two 60-day long
timeseries from the longer second run, LRa01 (data available at {\tt
idoc-CoRoT.ias.u-psud.fr}). The different fitters analysed this
spectrum independently, and we then compared the results.

 \section{Results}
 \label{sec:res}

The availability of the new longer CoRoT timeseries makes the mode
(degree $l$) identification significantly less ambiguous than it was
before.  The p-mode peaks are still observed to be very wide (several
$\mu$Hz), and the typical signal-to-noise ratio (SNR) (defined as the
ratio of the height of a mode to the level of the background around
the mode) is similar to that for the first, shorter CoRoT
run. However, the additional information provided by the longer second
run is sufficient to allow the modes to be tagged with far greater
confidence than was hitherto possible. There was a very good agreement
between the results of the fitters, with model B (which has an $l$=1
mode at 1755\,$\mu$Hz; see Table~\ref{tab:freq} for details) favoured
strongly over Model A. As part of the analysis for this
paper, we performed a Bayesian analysis \citep[using the method described by][]{benomar08,benomar09} 
to compare model A with model B, and this now
favours model B at a confidence level higher than 99\,\%. It is
important to add that this confidence level does depend to some extent
upon the hypothesis used for the modeling (e.g. on the a priori
constraints applied to the model parameters). Here, the 99\,\% level
was estimated for a model with fixed mode height ratios (see
Section~\ref{sec:method} above). When the fixed height ratio
constraint was relaxed so that individual heights were fitted to
modes of different $l$, we found that the confidence level for model B
dropped no lower than 95\,\%. (It is worth adding that the
\emph{average} fitted height ratio agrees, to within errors, with the
solar-like value we adopted as the fixed height ratio.)

We also established, using the Bayesian approach, that models that
include $l=2$ modes are strongly favoured over those which do not (at
a confidence level over 99.9\,\%). To check for evidence in the
Fourier spectrum of $l=3$ modes, we instead used a collapsed
spectrum. We could find no evidence for a significant excess of power
in the wings of the $l=1$ modes. Evidently, the SNR in the $l=3$ modes
is too low for them to be observed.

We make one final remark regarding the identification problem.
\citet{appourchaux08}, who favoured model A, made their choice based
solely on a comparison of the maximum likelihoods given by a classical
MLE analysis. When the fitting problem is non-trivial, this type of
analysis can converge on a subsidiary maximum (not the true, global
maximum), biasing any statistical comparison of two possible
models. In addition to MLE, we also applied for the work in this paper the
Markov Chain Monte-Carlo (MCMC) analysis, which circumvents the above
problem by giving a complete sampling of the parameter space. Both
MCMC and MLE now clearly favour Model B. The present results are due to the conjunction of two facts. First, the extra information provided by the new observations (adding 137 days to the first 60 days of observation). Second,  the confirmation given by the use of Bayesian analysis about a robust comparison of the associated probabilities to each model. All this ensures the identification without ambiguity, whatever the method used.


\begin{figure}
\begin{center}
\includegraphics[angle=90,height=6.4cm]{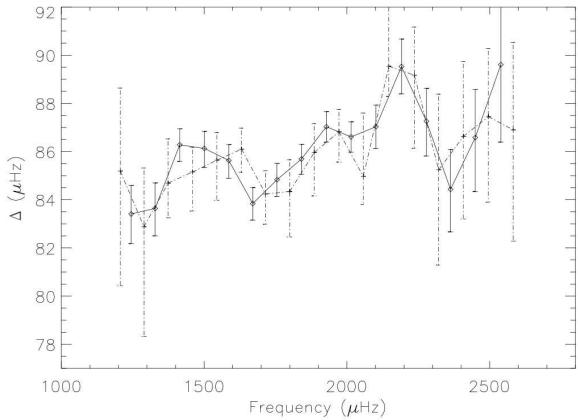}
\end{center}
 \caption{Large frequency separation computed from $l$=0 (crosses linked by a
 dashed line) and $l$=1 (diamonds linked by a solid line) modes.}
 \label{fig:largsep}
 \end{figure}


Most of the frequencies for model B returned by the different fitters
lie within a 1-$\sigma$ error interval (which corresponds to a level
of confidence of 68\%), and all do so within a 1.5-$\sigma$
interval. These error intervals are particularly small in the
frequency range $ \nu \in [1300,2000]\,\rm \mu Hz$ where the modes
have relatively large heights. Agreement is particularly good at $l$=1
($\sigma$\,$\sim$\,0.6$\mu$Hz; see Fig.\,9). This is because these modes are not affected by prominent nearby modes, as is
the case for the closely spaced $l=0$ and 2 modes (which show
significant overlap in frequency). The $l=2$ modes have lower
amplitudes than their $l=0$ neighbours and consequently have the
largest errors of any of the observed $l$ ($\sigma$\,$\sim$\,2$\mu$Hz,
see Fig.\,9).

\begin{figure}
\begin{center}
\includegraphics[angle=90,height=6.4cm]{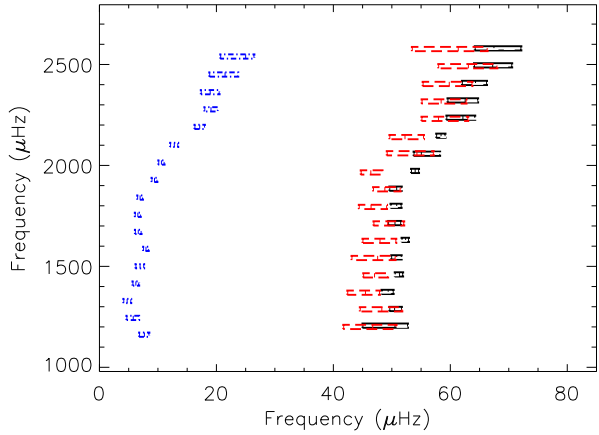}
\end{center}
\caption{Echelle diagram built from the fitted frequencies for a large separation of 85 $\mu$Hz. The error boxes
  indicate the 1$\sigma$ (68\% confidence) interval.}
\label{fig:diag_ech}
\end{figure}

\begin{figure}[htp]
\begin{center}
\includegraphics[height=7.6cm,width=6.4cm,angle=90.]{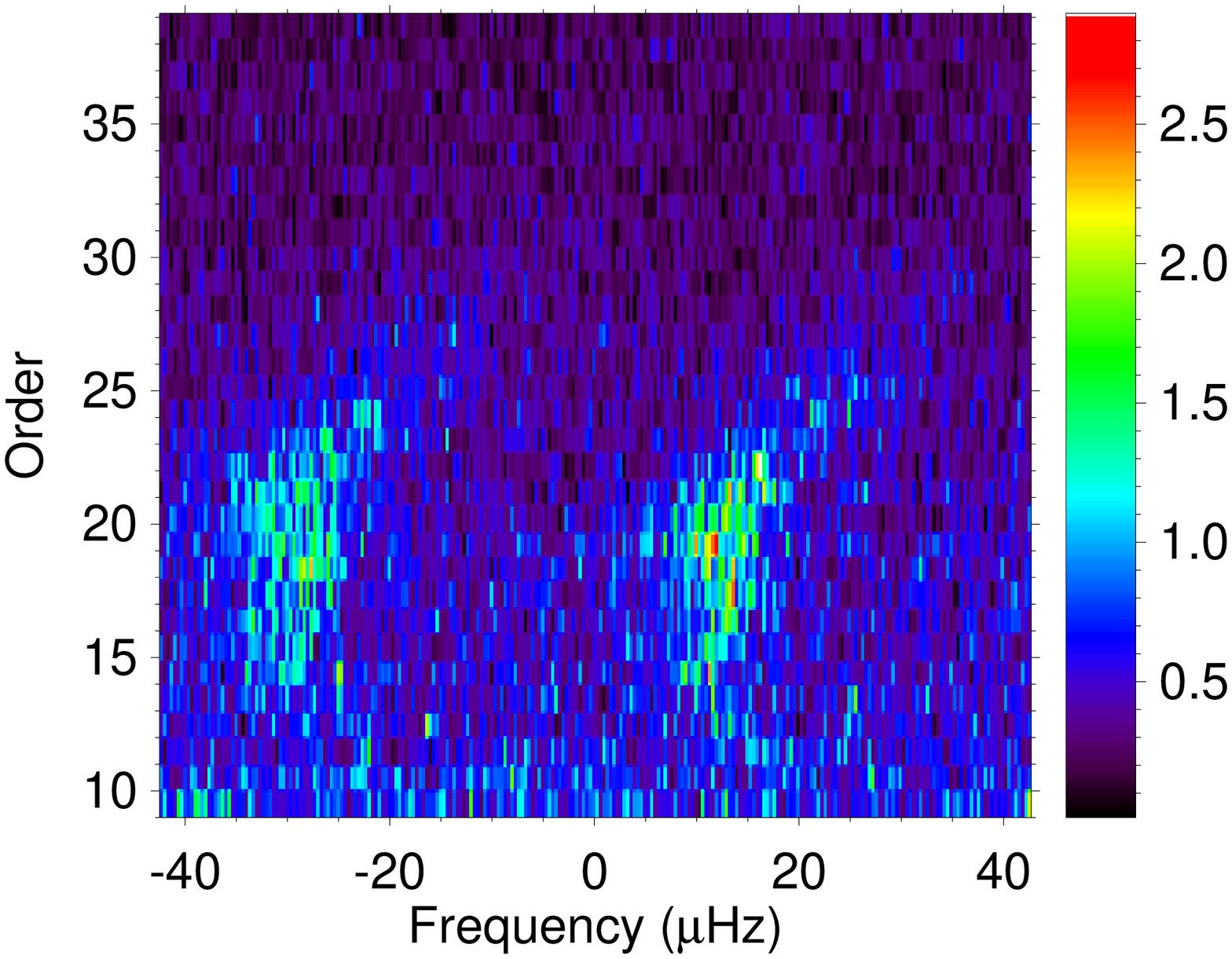}
\end{center}
\caption{Echelle diagram computed for a large separation of 85 $\mu$Hz for the mean power spectrum. The diagram is smoother to 0.8 $\mu$Hz. The identification of the ridges is far from evident. The most probable model correspond to $l=1$ for the left ridge and $l=0$ for the right ridge (same convention as in Fig.\,\ref{fig:diag_ech})}
\label{fig:diag_ech0}
\end{figure}

Table~\ref{tab:freq} lists Model B frequencies from one of the
analyses used here \citep[a Bayesian analysis coupled to MCMC
sampling; see for example][]{benomar09}. The \'echelle diagram of
these frequencies is shown in Fig.~\ref{fig:diag_ech}. From the most
reliable $l$=0 and $l$=1 modes, it is possible to estimate the large
frequency separations $\Delta_0$ and $\Delta_1$ and their variation
with frequency (see Fig.~\ref{fig:largsep}). The uncovered frequency
variation may be regarded as being significant, given the good
precision on the estimated frequencies. Any estimate of the frequency
difference $\delta_{02}$ between neighbouring $l$=0 and $l$=2 modes is
much less reliable because of the difficulty of fitting these modes
(see above). Thus, the only result that can be provided is the
average: $\langle \delta_{02}\rangle=4.7\mu$Hz.

\begin{figure}
\begin{center}
\includegraphics[angle=90,height=5.4cm]{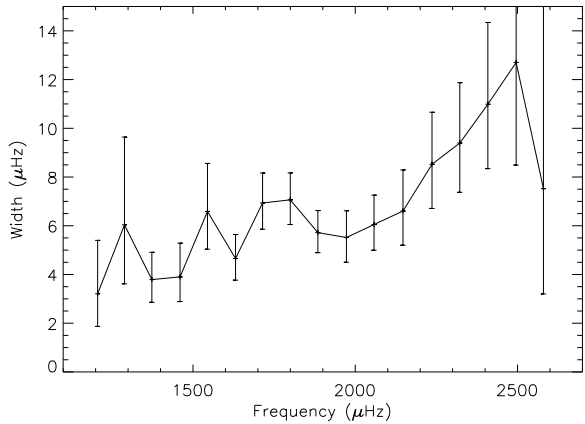}
\end{center}
\caption{Widths of the fitted modes, with 1$\sigma$ error bars}.
\label{fig:larg}
\end{figure}

\begin{figure}[ht]
\begin{center}
 \includegraphics[angle=90,height=6cm]{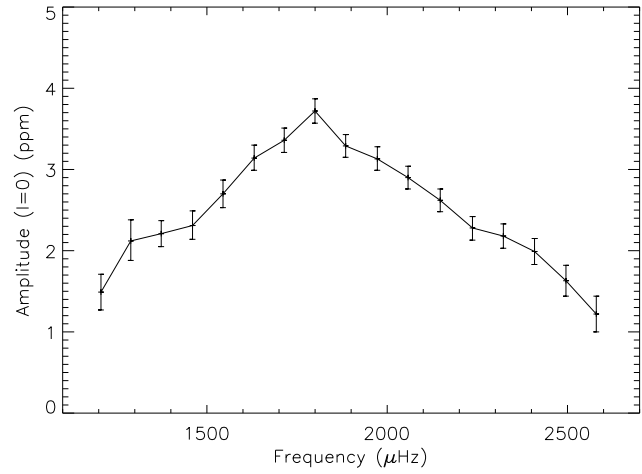}
\end{center}
\caption{Amplitude ($\sqrt{\pi H \Gamma}$ with $H$ the height and $\Gamma$ the width) of the $l\,=\,0$ fitted modes, with 1$\sigma$ error bars.}
\label{fig:amp}
\end{figure}

The widths of the modes, which are related to the damping, are also
listed in Table~\ref{tab:freq} and shown in Fig.~\ref{fig:larg}. This
figure illustrates the main difficulty of interpreting the spectrum:
the relatively high values of the widths. As on the Sun, an increase
in width with increasing frequency is visible, but the level of
precision prevents a more detailed analysis. The amplitudes of the
modes, which depend on the balance between damping and excitation, are
listed in Table~\ref{tab:freq}. The $l=0$ amplitudes are plotted in
Fig.~\ref{fig:amp}. The maximum is about $\sim$3.7\,ppm, which, while
higher than the maximum amplitude of low-$l$ solar oscillations, is
still lower than predicted by a scaling of amplitude on
$(L/M)^{\alpha}$ \citep[with $\alpha$\,$\simeq$\,0.7,
see][]{samadi07}.
Extraction of the rotational frequency splitting,
$\nu_s$, of the modes remains very difficult: the analyses of the
different fitters failed to converge on a unique solution. Estimated
values were in the range $3.5\mu$Hz\,$<\nu_s<$\,6.0$\mu$Hz, i.e., high
compared to the solar value, but as expected given the surface
rotation period estimated from the low-frequency part of the Fourier
spectrum ($P_{\rm rot} \simeq 3.4$ days). 
Finally, we were able to
extract robust values for the angle of inclination of the star. All
analyses converged on an angle of of $17^{\circ}\,^{+7}_{-9}$. This is
in agreement with an independent determination made using measurements of the stellar $v\sin i$, radius and period \citep{solano05,mosser09}. However, as mentioned earlier, such a small angle greatly favours the visibility of $m=0$ mode components, rendering the splitting measurement very difficult.

\begin{table}
 \begin{tabular}{|c|ccc|c|}
\hline
Degree & Frequency & 1$\sigma$ & 2$\sigma$ & Height/Noise \\
      $l$      & ($\mu$Hz) & interval & interval & ratio \\
\hline
 0 & 1206.25 & +2.23/-5.57& +3.92/-7.26 &0.6 \\
 0 & 1288.97 & +1.25/-0.90& +3.20/-1.97 &0.7 \\
 0 & 1373.34 & +0.93/-1.16& +1.70/-2.50 &1.3 \\
 0 & 1460.14 & +0.65/-0.80& +1.22/-2.15 &1.4 \\
 0 & 1544.69 & +0.85/-0.95& +1.76/-2.05 &1.2 \\
 0 & 1631.10 & +0.59/-0.69& +1.17/-1.59 &2.4 \\
 0 & 1714.49 & +1.06/-1.17& +2.00/-2.21 &1.9 \\
 0 & 1799.75 & +0.84/-1.07& +1.62/-2.62 &2.4 \\
 0 & 1884.82 & +0.88/-1.26& +1.63/-2.65 &2.4 \\
 0 & 1972.73 & +0.67/-0.71& +1.30/-1.59 &2.3 \\
 0 & 2057.82 & +3.22/-1.34& +4.72/-2.24 &1.9 \\
 0 & 2147.10 & +0.76/-0.83& +1.65/-1.99 &1.4 \\
 0 & 2236.46 & +2.11/-2.83& +3.85/-5.44 &0.9 \\
 0 & 2322.10 & +2.22/-3.03& +4.08/-5.64 &0.7 \\
 0 & 2408.56 & +1.97/-2.33& +3.75/-4.81 &0.5 \\
 0 & 2495.76 & +3.34/-3.21& +7.00/-6.23 &0.3 \\
 0 & 2579.85 & +4.70/-3.29& +7.61/-10.7 &0.3 \\
\hline
 1 & 1161.54 & +0.85/-0.91& +1.67/-2.25 &0.6 \\
 1 & 1244.63 & +1.02/-1.17& +2.39/-2.66 &0.7 \\
 1 & 1328.34 & +0.70/-0.65& +1.40/-1.26 &1.2 \\
 1 & 1414.93 & +0.57/-0.58& +1.22/-1.28 &1.4 \\
 1 & 1500.54 & +0.70/-0.78& +1.36/-1.62 &1.2 \\
 1 & 1586.62 & +0.48/-0.49& +0.98/-1.02 &2.3 \\
 1 & 1670.48 & +0.57/-0.58& +1.16/-1.23 &1.9 \\
 1 & 1755.30 & +0.51/-0.53& +1.06/-1.05 &2.3 \\
 1 & 1840.68 & +0.49/-0.50& +0.98/-1.02 &2.3 \\
 1 & 1928.13 & +0.50/-0.51& +1.00/-1.05 &2.3 \\
 1 & 2014.38 & +0.54/-0.54& +1.07/-1.09 &1.8 \\
 1 & 2101.58 & +0.67/-0.72& +1.42/-1.56 &1.4 \\
 1 & 2190.81 & +0.90/-0.90& +1.88/-1.92 &0.8 \\
 1 & 2277.89 & +1.16/-1.14& +2.33/-2.37 &0.7 \\
 1 & 2362.76 & +1.61/-1.61& +3.39/-3.48 &0.5 \\
 1 & 2450.35 & +2.26/-2.71& +4.58/-7.01 &0.3 \\
 1 & 2539.49 & +1.50/-4.35& +3.24/-11.2 &0.3 \\
\hline
 2 & 1199.91 & +4.08/-5.04& +8.77/-11.3 &0.6 \\
 2 & 1287.24 & +3.51/-3.77& +7.53/-8.65 &0.7 \\
 2 & 1369.60 & +2.48/-3.11& +5.25/-7.79 &1.3 \\
 2 & 1455.42 & +2.33/-1.94& +5.49/-4.07 &1.4 \\
 2 & 1541.54 & +3.07/-4.47& +5.94/-9.48 &1.2 \\
 2 & 1626.30 & +2.79/-3.00& +5.16/-6.17 &2.4 \\
 2 & 1712.67 & +2.83/-2.44& +5.23/-4.75 &1.9 \\
 2 & 1794.39 & +2.73/-2.10& +6.47/-3.91 &2.4 \\
 2 & 1881.83 & +2.70/-2.02& +5.27/-4.08 &2.4 \\
 2 & 1965.19 & +2.01/-1.74& +4.78/-3.55 &2.3 \\
 2 & 2060.22 & +2.88/-5.17& +4.57/-7.40 &1.9 \\
 2 & 2140.32 & +3.25/-2.67& +7.02/-5.08 &1.4 \\
 2 & 2230.68 & +5.10/-2.91& +9.99/-5.89 &0.9 \\
 2 & 2316.86 & +4.35/-3.35& +9.06/-8.24 &0.7 \\
 2 & 2403.52 & +3.97/-4.56& +7.77/-11.1 &0.5 \\
 2 & 2491.50 & +4.96/-5.16& +10.3/-11.2 &0.3 \\
 2 & 2576.51 & +5.03/-7.94& +9.92/-18.6 &0.3 \\

\hline
\end{tabular}
\caption{Frequencies of the fitted modes. The 1 and 2-$\sigma$
intervals correspond to confidence levels of 68\% and 95\%,
respectively.}
\label{tab:freq}
\end{table}

 \section{Conclusion}
 \label{sec:conc}

The two CoRoT observation runs on the star HD49933 -- the first 60-day
run, and the more recent longer run -- have now provided enough data
to resolve the identification of modes in the oscillation spectrum at
a very high confidence level. This identification relied on several
independent analyses.  The new data have also allowed us to improve the
precision in the mode parameters, with fractional improvements being
in the range from 40\,\% to 70\,\% depending on the parameter. It is
now possible to determine precise mode frequencies for $l$=0 and 1
modes. The $l$=2 mode parameters are more difficult to estimate
because of the overlap with the stronger, neighbouring $l$=0
modes. The widths and amplitudes of the modes are well determined, as
is the inclination angle of the star. The rotational frequency
splitting remains the only mode parameter that is poorly
constrained. However, it is clearly much higher than the solar value,
and similar or larger in size to the inverse of the surface rotation
period. In summary, this new information on the acoustic oscillations
of HD49933 now opens the possibility for detailed seismic modeling of
the star.

\begin{acknowledgements}

W.J.C. and Y.E. wish to thank the UK Science and Technology Facilities
Council (STFC) for support under grant ST/F00204/1. I.W.R. and
G.A.V. also thank STFC, for support under grant
PP/E001793/1. J.B. acknowledges support through the ANR project
Siroco. We thank J. Leibacher for very useful comments.

\end{acknowledgements}

\bibliographystyle{aa}
\bibliography{13111}

\clearpage
\onecolumn{
\begin{center}
\begin{Large}Online Material\end{Large}
\end{center}}

\begin{table}[ht]
\begin{center}
 \begin{tabular}{|c|cc|cc|}
\hline
Degree & Amplitude & 1$\sigma$ & Width & 1$\sigma$ \\
 $l$   & (ppm)      &  interval & ($\mu$Hz) &  interval \\
\hline
0 & 1.49 & +0.22/-0.22 & 3.21 & +2.19/-1.34 \\
0 & 2.12 & +0.26/-0.24 & 6.04 & +3.60/-2.42 \\
0 & 2.21 & +0.16/-0.16 & 3.79 & +1.12/-0.93 \\
0 & 2.31 & +0.18/-0.17 & 3.90 & +1.39/-1.01 \\
0 & 2.70 & +0.17/-0.17 & 6.59 & +1.97/-1.55 \\
0 & 3.14 & +0.16/-0.15 & 4.66 & +0.99/-0.89 \\
0 & 3.36 & +0.15/-0.15 & 6.94 & +1.23/-1.08 \\
0 & 3.72 & +0.15/-0.15 & 7.06 & +1.11/-1.01 \\
0 & 3.29 & +0.14/-0.14 & 5.72 & +0.90/-0.82 \\
0 & 3.13 & +0.15/-0.14 & 5.52 & +1.10/-1.02 \\
0 & 2.90 & +0.14/-0.14 & 6.05 & +1.21/-1.05 \\
0 & 2.62 & +0.14/-0.14 & 6.61 & +1.69/-1.40 \\
0 & 2.28 & +0.14/-0.15 & 8.52 & +2.14/-1.81 \\
0 & 2.18 & +0.15/-0.15 & 9.40 & +2.48/-2.02 \\
0 & 1.99 & +0.16/-0.16 & 11.0 & +3.36/-2.64 \\
0 & 1.63 & +0.19/-0.19 & 12.7 & +6.71/-4.21 \\
0 & 1.22 & +0.22/-0.22 & 7.52 & +8.72/-4.32 \\
\hline
1 & 1.82 & +0.26/-0.28 & 3.21 & +2.19/-1.34 \\
1 & 2.58 & +0.31/-0.30 & 6.04 & +3.60/-2.42 \\
1 & 2.70 & +0.19/-0.20 & 3.79 & +1.12/-0.93 \\
1 & 2.82 & +0.22/-0.21 & 3.90 & +1.39/-1.01 \\
1 & 3.30 & +0.21/-0.21 & 6.59 & +1.97/-1.55 \\
1 & 3.83 & +0.19/-0.19 & 4.66 & +0.99/-0.89 \\
1 & 4.10 & +0.19/-0.19 & 6.94 & +1.23/-1.08 \\
1 & 4.54 & +0.18/-0.18 & 7.06 & +1.11/-1.01 \\
1 & 4.01 & +0.17/-0.17 & 5.72 & +0.90/-0.82 \\
1 & 3.82 & +0.18/-0.17 & 5.52 & +1.10/-1.02 \\
1 & 3.54 & +0.16/-0.17 & 6.05 & +1.21/-1.05 \\
1 & 3.19 & +0.18/-0.18 & 6.61 & +1.69/-1.40 \\
1 & 2.78 & +0.17/-0.18 & 8.52 & +2.14/-1.81 \\
1 & 2.66 & +0.18/-0.18 & 9.40 & +2.48/-2.02 \\
1 & 2.43 & +0.19/-0.20 & 11.0 & +3.36/-2.64 \\
1 & 1.99 & +0.23/-0.23 & 12.7 & +6.71/-4.21 \\
1 & 1.49 & +0.28/-0.27 & 7.52 & +8.72/-4.32 \\
\hline
2 & 1.09 & +0.16/-0.16 & 3.21 & +2.19/-1.34 \\
2 & 1.54 & +0.19/-0.18 & 6.04 & +3.60/-2.42 \\
2 & 1.61 & +0.12/-0.12 & 3.79 & +1.12/-0.93 \\
2 & 1.68 & +0.13/-0.12 & 3.90 & +1.39/-1.01 \\
2 & 1.97 & +0.13/-0.12 & 6.59 & +1.97/-1.55 \\
2 & 2.28 & +0.12/-0.11 & 4.66 & +0.99/-0.89 \\
2 & 2.44 & +0.11/-0.11 & 6.94 & +1.23/-1.08 \\
2 & 2.71 & +0.11/-0.11 & 7.06 & +1.11/-1.01 \\
2 & 2.39 & +0.10/-0.10 & 5.72 & +0.90/-0.82 \\
2 & 2.28 & +0.11/-0.10 & 5.52 & +1.10/-1.02 \\
2 & 2.11 & +0.10/-0.10 & 6.05 & +1.21/-1.05 \\
2 & 1.91 & +0.10/-0.10 & 6.61 & +1.69/-1.40 \\
2 & 1.66 & +0.10/-0.11 & 8.52 & +2.14/-1.81 \\
2 & 1.59 & +0.11/-0.11 & 9.40 & +2.48/-2.02 \\
2 & 1.45 & +0.12/-0.12 & 11.0 & +3.36/-2.64 \\
2 & 1.19 & +0.14/-0.14 & 12.7 & +6.71/-4.21 \\
2 & 0.89 & +0.16/-0.16 & 7.52 & +8.72/-4.32 \\
\hline
\end{tabular}
\caption{Amplitudes and widths of the fitted modes. The 1-$\sigma$
intervals correspond to confidence levels of 68\%.}
\label{tab:amp_width}
\end{center}
\end{table}

\begin{figure}\label{fig:errfreq}
\begin{center}
\includegraphics[angle=90,height=6cm]{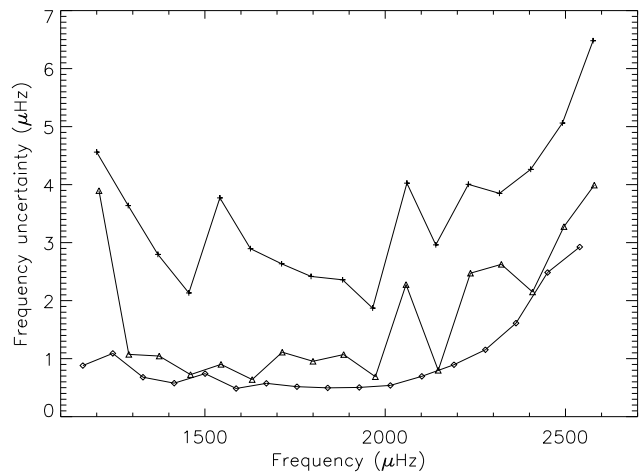}
\caption{Mean (upper and lower) 1$\sigma$ error bars for the
frequencies of $l$=0 (triangles), $l$=1 (diamonds) and $l$=2 (crosses)
modes, showing the decreasing precision on the frequencies from $l$=1
to $l$=2 modes.}
\end{center}

\end{figure}

\end{document}